\documentclass[10pt, two column, twoside]{IEEEtran}

\usepackage{amssymb}
\usepackage{amsmath}
\usepackage[lined,boxed,commentsnumbered, ruled]{algorithm2e}
\usepackage{mathrsfs}
\usepackage{algorithmic}
\usepackage{bm}

\usepackage{tikz}
\usetikzlibrary{arrows}
\usepackage{subfigure}
\usepackage{graphicx,booktabs,multirow}

\definecolor{colorhkust}{RGB}{20,43,140}
\definecolor{colortsinghua}{RGB}{116,52,129}
\definecolor{color1}{RGB}{128,0,0}

\newcommand{\trace}{{\rm Tr}}
\newcommand{\Sym}{{\mathrm{Sym}}}
\newcommand{\grad}{\mathrm{grad}}
\newcommand{\rc}{\nabla}
\newcommand{\D}{\mathrm{D}}
\newcommand{\GL}[1]{{\mathrm{GL}({#1})}}
\newcommand{\mat}[1]{{\bm #1}}
\newcommand{\rank}{\mathrm{rank}}
\newcommand{\subject}{\sf{subject\  to}}
\newcommand{\hess}{\mathrm{Hess}}

\newcommand{\changeBM}[1]{#1}
\newcommand{\changeYS}[1]{#1}

\begin{document}


\title{A Sparse and Low-Rank Optimization  Framework for Index Coding via Riemannian Optimization
}
\author{\IEEEauthorblockN{Yuanming Shi$^{\alpha}$ and Bamdev Mishra$^{\beta}$}\\
\IEEEauthorblockA{$^{\alpha}$School of Information Science and Technology, ShanghaiTech University, Shanghai, China\\
                           E-mail: shiym@shanghaitech.edu.cn}\\
\IEEEauthorblockA{$^{\beta}$Amazon Development Centre India, Bangalore, Karnataka 560055, India  \\
                           E-mail: bamdevm@amazon.com}
                           }

\maketitle
\IEEEpeerreviewmaketitle


\maketitle

\begin{abstract}
Side information provides a pivotal role for message delivery in many communication scenarios to accommodate increasingly large data sets, e.g., caching networks. Although index coding provides a fundamental modeling framework to exploit the benefits of side information, the index coding problem itself still remains open and only a few instances have been solved. In this paper, we propose a novel sparse and low-rank optimization modeling framework for the index coding problem to characterize the tradeoff between the amount of side information and the achievable data rate. Specifically, sparsity of the model measures the amount of side information, while low-rankness represents the achievable data rate. \changeBM{The resulting sparse and low-rank optimization problem has non-convex sparsity inducing objective and non-convex rank constraint.} To address the coupled challenges in objective and constraint, we propose a novel Riemannian optimization framework by exploiting the quotient manifold geometry of fixed-rank matrices, accompanied by \changeBM{a} smooth sparsity inducing surrogate. Simulation results demonstrate the appealing sparsity and low-rankness tradeoff in the proposed model, thereby revealing the tradeoff between the amount of side information and the achievable data rate in the index coding problem.

\end{abstract}


\IEEEpeerreviewmaketitle

\section{Introduction}
With the dramatic increase of smart mobile devices, as well as diversified services and applications, we are in the era of data deluge \cite{Ding_CMagBigData}. Meanwhile, with the emerging applications empowered by the Internet of Things
(IoT) and Tactile Internet, massive devices will need to get connected, which calls for ultra-low latency, high availability, reliability and security communications \cite{Fettweis_JSAC2016}. However, with the low latency and high data rate requirements, the communication systems are placed under tremendous pressure to accommodate increasingly large data sets and to efficiently deliver the content. To resolve the big data challenge in communication networks, side information plays a pivotal role for both the wired and wireless communication links to deliver messages to users \cite{Niesen_TIT2014Caching,Jafar_TIT2014indexcoding}. That is, users can access to the messages as the side information that requested by other users. For instance, this scenario arises in the cache enabled fog radio access networks (Fog-RAN) \cite{Yuanming_WCMLargeCVX}. In this network architecture, the content can be stored in the caches or other storage elements, e.g., the fog data center, radio access points and the mobile devices. The cached content may be requested by other users or in the further, thereby providing side information for message delivery in wired and wireless communications \cite{Niesen_TIT2014Caching,Caire_TIT2016D2Dcaching}.

Index coding  provides a powerful framework to model the communication scenarios with side information \cite{Jafar_TIT2014indexcoding}. Although it has been shown that the index coding problem is related to many challenging problems (e.g., distributed storage, topological interference management \cite{Yuanming_2016LRMCTWC} and network coding \cite{Langberg_TIT2015}), the index coding problem itself remains open. Most of works on index coding focus on how to exploit the fixed side information, thereby designing efficient message delivering strategies, e.g., the interference alinement approach \cite{ Jafar_TIT2014indexcoding}. In particular, in caching networks \cite{Niesen_TIT2014Caching}, the side information (i.e., massage placement) can be designed, followed by the message delivery. However, the amount of the side information in caching networks is limited by the storage capacity in caching networks.       

In this paper, we put forth a different viewpoint on the index coding problem by investigating the fundamental tradeoff between the amount of the side information and the achievable data rate. That is, the higher data rate comes from at the price of high storage size, yielding more side information. To achieve this goal, we propose a novel sparse and low-rank optimization framework to minimize the amount of side information to meet a data rate requirement. Specifically, the sparsity of this model represents the amount of side information, while the low-rankness of this model represents the number of channel uses, i.e., blocklength, which equals the inverse of the achievable data rate. Although the sparse and low-rank models have recently been well-studied in signal processing and machine learning \cite{Hassibi_TIT2015sl}, the presented model for index coding is novel and can help reveal the fundamental tradeoff between the amount of side information and the achievable data rate.

Unfortunately, the resulting sparse and low-rank optimization problem \changeBM{raises a} unique challenge due to a non-convex objective function \changeBM{($\ell_0$)} and non-convex constraint \changeBM{(rank)}. Although the convex relaxation approach based on convex surrogates -- $\ell_1$-norm and nuclear norm -- can provide polynomial time complexity algorithms \cite{Hassibi_TIT2015sl}, this approach is inapplicable in our problem as it always return the identity matrix. Another approach is based on alternating minimization by factorizing a fixed-rank matrix \cite{Wotao_2012solvingLR}, accompanied with $\ell_1$-norm relaxation. \changeYS{However, this approach fails to yield good performance by inducing a less sparse solution and is computationally expensive using the off-the-shelf parser/solver CVX \cite{cvx}.} 

To address the limitations of the above methods, we propose a Riemannian optimization algorithm \cite{Absil_2009optimizationonManifolds} to solve the resulting sparse and low-rank optimization problem. In particular, by exploiting the quotient manifold geometry of fixed-rank matrices \cite{Mishra_2014fixedrank}, the Riemannian optimization algorithm was proposed to solve the low-rank matrix completion problem for topological interference management \cite{Yuanming_2016LRMCTWC}. However, this algorithm can not be applied in our problem due to the additional affine constraint preserving the desired signals and the non-convex sparsity inducing objective. We thus propose a smooth  sparsity inducing surrogate and regularize the affine constraint as a smooth least-squares term. The second-order trust-region method \cite{Absil_2009optimizationonManifolds} is further applied to the resulting optimization problem with smooth objective over fixed-rank manifold constraint. The proposed algorithm, \changeYS{which \changeBM{is} implemented in the manifold optimization \changeBM{toolbox} Manopt \cite{manopt}}, outperforms \changeBM{the alternating minimization algorithm} in terms of {\changeYS{implementation complexity}} and performance. Simulation results demonstrate the appealing tradeoff between the sparsity and low-rankness of the model, thereby revealing the tradeoff between the amount of side information and the achievable data rate.          

\section{Problem Statement}
We consider the communication networks (e.g., caching network \cite{Niesen_TIT2014Caching}) with side information to help message delivery. To investigate the tradeoff between the amount of side information and the achievable data rate, we introduce an index coding modeling framework for communications with side information \cite{Jafar_TIT2014indexcoding}. Specifically, we consider a multiple unicast index coding problem consists of a set of $K$ independent messages $W_1,W_2,\dots, W_K$, and a set of $K$ destination nodes.
The $i$-th destination desires message $W_i$ with side information index
as $\mathcal{V}_i$ and $i\notin\mathcal{V}_i$. 

Let $\mathcal{S}$ be the choice of a finite alphabet. The coding function $f$ for all the messages is given by
\setlength\arraycolsep{2pt}
$f(W_1,W_2,\dots, W_K)={\bm{z}}$,
where $\bm{z}\in\mathcal{S}^N$ is the sequence of symbols transmitted over
$N$ channel uses. Here, each message $W_i$ is a
random variable uniformly distributed over the set $W_i\in\{1,2,\dots, |\mathcal{S}|^{N
R_i}\}$ with $|\mathcal{S}|^{NR_i}$ as an integer. At destination $i$, the decoding function $g_i$ for the desired message $W_i$ is given by
$g_i(\bm{z},\mathcal{V}_i)=\hat W_i$.
The probability of decoding error is
given by
$p_e=1-{\rm{Pr}}\{\hat W_i=W_i, \forall i\}$.

Define the above coding scheme as $(\mathcal{S}, N, (R_1,\dots, R_K))$. If for every $\epsilon,\delta>0$, for some $\mathcal{S}$ and $N$, there exists a coding scheme $(\mathcal{S}, N, (\bar{R}_1,\bar{R}_2,\dots, \bar R_K))$, such that $\bar{R}_i\ge R_i-\delta, \forall i$, and the error probability $P_e\le \epsilon$, then the rate tuple $(R_1,R_2,\dots, R_K)\in\mathbb{R}_{+}^K$ is said to be achievable. Note that the index coding capacity does not dependent on the field specification \cite{Jafar_TIT2014indexcoding}. In this paper, the achievable scheme is restricted to the real field $\mathbb{R}$ for linear coding schemes design to construct index codes over real field.

\subsection{Scalar Linear  Index Coding Scheme}
Consider a scalar linear index coding scheme, which sends one symbol for
each message over $N$ channel uses. Let $\bm{v}_i\in\mathbb{R}^N$ and $\bm{u}_i\in\mathbb{R}^N$ be the precoding vector and the decoding vector, respectively. The transmitted symbol
sequence $\bm{z}\in\mathbb{R}^{N\times1}$ over $N$ channel uses in a linear
coding scheme is given by
$\bm{z}=\sum_{i=1}^K\bm{v}_is_i$,
where $s_i$ is one symbol from $\mathbb{R}$ representing $W_i$. The decoding
operation for message $W_k$ at destination $k$ is given by
\begin{eqnarray}
\hat{s}_k=\left({\bm{u}}_k^{T}{\bm{v}}_k\right)^{-1}{\bm{u}}_k^{T}\left({\bm{z}}-\sum\nolimits_{i\in\mathcal{V}_k}{\bm{v}}_is_i\right).
\end{eqnarray}

The above decoding operation is achieved by the following interference alignment
condition \cite{Jafar_TIT2014indexcoding,Yuanming_2016LRMCTWC}:
\begin{eqnarray}
\label{c1}
{\bm{u}}_k^{T}{\bm{v}}_k&\ne& 0, \forall k=1,\dots, K\\
\label{c2}
{\bm{u}}_k^{T}{\bm{v}}_i&=&0, \forall i\ne k, i\notin\mathcal{V}_k.
\end{eqnarray}
If the above interference alignment conditions (\ref{c1}) and (\ref{c2})
are satisfied over $N$ channel uses, the following data rate vector
$\mathcal{R}=\left({1\over{N}},{1\over{N}},\dots, {1\over{N}}\right)$,
can be achieved \cite{Jafar_TIT2014indexcoding,Yuanming_2016LRMCTWC}. Therefore, the achievable sum data rate is given by $K/N$.

\subsection{Storage Size and Data Rate Tradeoff}

\changeYS{We shall give an example of  caching network to illustrate the amount of side information and achievable data rate.} Specifically, we assume that the file library is a set of $K$ messages
$\{W_1,\dots, W_K\}$, where each has entropy $F$ bits. Given the side information $\mathcal{V}_i$'s after content placement, the amount
of side information is  $\sum_{i=1}^K|\mathcal{V}_i|F$ bits, which measures
the total storage size. In this case, we characterize the tradeoff between the following two important metrics:
\begin{itemize}
\item The amount
of side information: $s:=\sum_{i=1}^K|\mathcal{V}_i|$ (normalized by $F$);
\item The achievable data rate: $r:=1/N$ (normalized by $K$).
\end{itemize}
In general, the more side information is available, the higher data rate can be achieved.

The index coding problem has shown to be related to the network coding problem
\cite{Langberg_TIT2015}, topological interference management problem \cite{Yuanming_2016LRMCTWC},
caching problem \cite{Niesen_TIT2014Caching}, as well as distributed storage
problem. This paper thus can provide principles for all these important design problems by characterizing the tradeoff between the amount of the side information and the achievable data rate.

\section{A Sparse and Low-Rank Optimization Framework for Index Coding}
In this section, we \changeBM{propose} a unified sparse and low-rank modeling framework to investigate the tradeoffs between  the amount of side information $\sum_i|\mathcal{V}_i|$ and the achievable data rate $1/N$ in the index coding problem. This is achieved by rewriting the interference alignment conditions (\ref{c1}) and (\ref{c2}) into a sparse minimization problem with a fixed-rank constraint and an affine constraint.

\subsection{Sparse and Low-Rank Modeling Framework}
\begin{figure}[t]
  \centering
  \includegraphics[width=0.95\columnwidth]{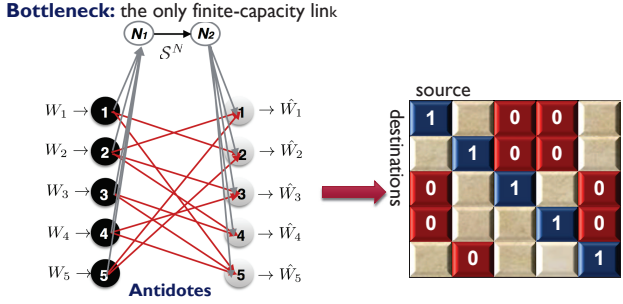}
 \caption{\changeYS{(a) The index coding problem with only one finite capacity link and all the other links having infinite capacity. The side information is given by $\mathcal{V}_1=\{2, 5\}, \mathcal{V}_2=\{1, 5\}, \mathcal{V}_3=\{2, 4\}, \mathcal{V}_4=\{2, 3\},\mathcal{V}_5=\{1,3,4\}$. (b) The associated
 incomplete matrix representing the interference alignment conditions (\ref{c1}) and (\ref{c2}).}}
 \label{TIM_example}
 \end{figure}

Let $X_{ij}={\bm{u}}_i^{T}{\bm{v}}_j, \forall i,j=1,\dots,
K$. Define the $K\times K$ matrix $\bm{X}=[X_{ij}]$, we have the rank of matrix $\bm{X}$ as
${\rm{rank}}(\bm{X})=N$. 
The  achievable data rate (normalized by $K$) is given by
\begin{eqnarray}
\label{rate}
r={1/{\rm{rank}}(\bm{X})}.
\end{eqnarray}
\changeBM{Additionally,} the sparsity of the matrix $\bm{X}$ is given by
$\|\bm{X}\|_0=\sum_{i=1}^{K}|\mathcal{V}_i|+K$. 
\changeBM{Finally,} the amount of side information (normalized by $F$) is given by
\begin{eqnarray}
\label{side}
s=(\|\bm{X}\|_0-K).
\end{eqnarray}
An example of the sparsity and low-rankness of the matrix $\bm{X}$ for the index coding problem is shown in \changeBM{Fig. \ref{TIM_example}}. \changeYS{In this case, the amount of side information is given by $s=\sum_{i}|\mathcal{V}_i|=11$, which equals $(\|\bm{X}\|_0-K)=11$ by assuming that the unknown entries in the associated incomplete matrix are non-zero.}

From (\ref{rate}) and (\ref{side}), we can see that, to characterize the tradeoff between the amount of side information (i.e., storage size) and the achievable data rate, \changeBM{it} is equivalent to characterize the tradeoff between the sparsity and low-rankness of the modeling matrix $\bm{X}$. Specifically, we propose to solve the following sparse and low-rank optimization problem:
\setlength\arraycolsep{1.5pt}
\begin{eqnarray}
\mathscr{P}: \mathop {\sf{minimize}}\limits_{{\bm X} \in \mathbb{R}^{K\times K}}&& \|{\bm{X}}\|_0\nonumber\\
\subject&&X_{ii}=1, \forall i=1,\dots, K\\
&&{\rm{rank}}({\bm{X}})=r, \nonumber
\end{eqnarray} 
where $r$ is a fixed rank value of matrix $\bm{X}$. 
By solving a sequence of the optimization problem $\mathscr{P}$ via varying \changeBM{$r$ from $1$ to $K$}, we can reveal the tradeoff between the sparsity and low-rankness of matrix $\bm{X}$.  

\subsection{Problem Analysis}
The widely used $\ell_1$-norm and nuclear-norm relaxation method provides a computationally tractable algorithm for the sparse and low-rank optimization as follows \cite{Hassibi_TIT2015sl}:
\begin{eqnarray}
\label{slr}
 \mathop {\sf{minimize}}\limits_{{\bm X} \in \mathbb{R}^{K\times
K}}&& \|{\bm{X}}\|_1+\lambda\|\bm{X}\|_*\nonumber\\
\subject&&X_{ii}=1, \forall i=1,\dots, K,
\end{eqnarray}
where $\lambda\ge 0$ is a regularized parameter, $\|\bm{X}\|_1:= \sum_{ij} |X_{ij}|$, and \changeBM{$\|\bm{X}\|_*$ is the nuclear norm of $\mat X$, i.e., it is defined as the summation of the singular values of $\mat X$.} \changeBM{$\|\bm{X}\|_1$ and $\|\bm{X}\|_*$ are popular convex surrogates of $\|{\bm X} \|_0$ and the rank constraint, respectively}. Unfortunately, \changeBM{since} $\|\bm{X}\|_*\ge|{\rm{Tr}}(\bm{X})|$ \cite{Yuanming_2016LRMCTWC} and $\|\bm{X}\|_1\ge K$, \changeBM{the} problem (\ref{slr}) always returns $\bm{X}=\bm{I}_K$ as solution, \changeBM{which clearly is not low rank}. 


Another \changeBM{approach} is based on alternating minimization by factorizing the rank-$r$ matrix $\bm{X}$ as $\bm{U}{\bm{V}}^T$, where ${\bm{U}}\in\mathbb{R}^{K\times r}$ and $\bm{V}\in\mathbb{R}^{K\times r}$ \changeBM{are full column rank matrices}. Consequently, problem $\mathscr{P}$ is further relaxed as follows:
\begin{equation}\label{l1norm_factorized}
\begin{array}{lll}
\mathop{\sf{minimize}}\limits_{{\bm U}, {\bm V} \in \mathbb{R}^{K\times r} } & & \|{\bm{U}}{\bm{V}}^{T}\|_1\\
\subject & & [\bm{U}\bm{V}^{T}]_{ii}=1, \forall i=1,\dots, K,
\end{array} 
\end{equation}    
where $[\cdot]_{ij}$ denotes the $(i,j)$-entry of a matrix. The alternating minimization algorithm for problem (\ref{l1norm_factorized}) consists of alternatively solving for $\bm{U}$ and $\bm{V}$ while fixing the other factor. However, the alternating minimization algorithm fails to exploit the second-order information to improve the performance, i.e., enhance sparsity in matrix $\bm{X}$. 

In this paper, in order to enhance sparsity via exploiting the second-order information, we propose a Riemannian optimization algorithm to \changeBM{approximately} solve problem $\mathscr{P}$.

\section{Riemannian Optimization Algorithm}
In this section, we propose a Riemannian optimization algorithm to solve problem $\mathscr{P}$. Specifically, the $\ell_0$-norm is relaxed to the $\ell_1$-norm, resulting in the optimization problem:
\begin{eqnarray}
\label{l1norm}
\mathop {\sf{minimize}}\limits_{{\bm X} \in \mathbb{R}^{K\times K}}&& \|{\bm{X}}\|_1\nonumber\\
\subject &&X_{ii}=1, \forall i=1,\dots, K\\
&&{\rm{rank}}({\bm{X}})=r.\nonumber
\end{eqnarray} 
\changeBM{However, the intersection of rank constraint and the affine constraint is challenging to characterize}. We, \changeBM{therefore}, propose to solve (\ref{l1norm}) in two steps. In the first step, we find a good sparsity pattern by considering a regularized version of (\ref{l1norm}). In the second step, we refine the estimate obtained in the first step. In both of these steps, the underlying step is an optimization problem over the set of fixed-rank matrices. The overall algorithm is presented in Table \ref{tab:overall_algorithm}.

\subsection{Finding Sparsity Pattern}
In the first step, we reformulate problem (\ref{l1norm}) as the \emph{regularized}
problem:
\begin{equation}\label{eq:regularized_formulation}
\begin{array}{lll}
\mathop {\sf{minimize}}\limits_{{\bm X} \in \mathbb{R}^{K\times K}} &&   \displaystyle\frac{1}{2}\sum\limits_{i
=1}^K ({X}_{ii} - 1)^2 + \rho  \sum\limits_{ij}\left({X}_{ij}^2 + \epsilon
^2\right)^{1/2} \\
\subject &&  \rank(\bm{X}) = r, \\
\end{array}
\end{equation}
where $\rho\ge 0$ is the regularization parameter and $\epsilon$ is the parameter
that approximates $|{X}_{ij}|$ with the smooth term $\left({X}_{ij}^2 + \epsilon ^2\right)^{1/2}$ that makes the objective function \emph{differentiable}. A very small $\epsilon$ leads to ill-conditioning of the objective function in (\ref{eq:regularized_formulation}). Similarly, a larger $\rho$ induces more sparsity in $\bm{X}$. Since we intend to obtain the sparsity pattern of the optimal $\bm X$, we set $\epsilon$ to a high value, \changeBM{e.g., $0.01$}, to make the problem (\ref{eq:regularized_formulation}) well conditioned. 

If $\bm{X}_{\rm opt}=[{{X}_{\rm opt}}_{ij}]$ is the solution of (\ref{eq:regularized_formulation}), then the sparsity pattern matrix $\bm{P} =[P_{ij}]$ is of size $K\times K$ such that ${P}_{ij} = 1$ if ${{X}_{\rm opt}}_{ij} > \epsilon$ and ${P}_{ij} = 0$ otherwise.



\subsection{Refining the Estimate}
Once the sparsity pattern $\bm P$ is determined by solving (\ref{eq:regularized_formulation}), the \emph{refining step} translates into solving a rank-constrained \emph{matrix completion} problem. \changeBM{To see this, note} that we know the positions of zeros in the solution matrix (from $\bm{P}$) and that the diagonal entries are all $1$s. \changeBM{Consequently,} computing the entries at other positions is \emph{equivalent} to the problem
 \begin{equation}\label{eq:refining}
\begin{array}{llll}
\mathop{\sf{minimize}}\limits_{{\bm X} \in \mathbb{R}^{n\times n}} &&   \displaystyle\frac{1}{2}\sum\limits_{i =1} ^K
({X}_{ii} - 1)^2 +  \displaystyle\frac{1}{2} ||({\bm P}.*\bm{X}) - {\bm X}||_F^2
\\
\subject & & \rank(\bm{X}) = r, \\
\end{array}
\end{equation}
where $\|\cdot \|_F$ is the \emph{Frobenius} norm of a matrix and ${\bm P}.*\bm{X}$ is the element-wise multiplication of the matrices $\bm P$ and $\mat X$. Additionally, the algorithm for (\ref{eq:refining}) is initialized from $\bm{X}_{\rm opt}$, \changeBM{which is} the solution of (\ref{eq:regularized_formulation}).

\begin{table}[t]
\caption{Riemannian optimization Algorithm for $\mathscr{P}$.}
\label{tab:overall_algorithm} 
\begin{center} 
\begin{tabular}{ |p{8.2cm}| }
\hline


\begin{itemize}
\setlength\itemsep{0.3em}
\item Finding sparsity partition: we solve the \emph{regularized} formulation (\ref{eq:regularized_formulation}) to identify a good \emph{sparsity} pattern $\bm P$, which is a binary matrix of size $K \times K$ with $1$s at \changeBM{non-zero} positions and $0$s at \changeBM{zero positions}.

\item Refining: once the sparsity pattern $\bm P$ is determined, we solve the matrix completion problem (\ref{eq:refining}) with rank constraint to refine the estimate obtained from the regularized formulation solution. 

\item Both (\ref{eq:regularized_formulation}) and (\ref{eq:refining}) are solved with a Riemannian trust-region algorithm on the set of fixed-rank matrices. 
\end{itemize}
\\
\hline
\end{tabular}
\end{center} 
\end{table}

\subsection{Fixed-Rank Riemannian Manifold Optimization}\label{sec:fixed_rank}
The optimization problems (\ref{eq:regularized_formulation}) and (\ref{eq:refining}) are regularized \emph{least-square} optimization problems over the set of fixed-rank matrices. A rank-$r$ matrix $\mat{X} \in \mathbb{R}^{K \times K}$ is factorized as $\mat{X} = \mat{U} \mat{V}^T$, where $\mat{U} \in \mathbb{R} ^{K \times r}$ and $\mat{V} \in \mathbb{R}^{K
\times r}$ are full column-rank matrices. Such a factorization, however, is not unique as $\mat{X}$ remains unchanged under the transformation of the factors
\begin{equation}\label{eq:symmetry_gh}
(\mat{U},\mat{V})\mapsto (\mat{U}\mat{M}^{-1},\mat{V}\mat{M}^{T}),
\end{equation}
for all non-singular matrices $\mat{M} \in \GL{r}$, the set of $r \times r$ non-singular matrices. \changeBM{Equivalently, $\mat{X} = \mat{U} \mat{V}^T = \mat{U} \mat{M}^{-1}  (\mat{V}   \mat{M}^T)^T$ for all non-singular matrices $\mat M$.} As a result, the local minima of an objective function \changeBM{parameterized with $\mat U$ and $\mat V$} are not isolated on $\mathbb{R} ^{K \times r} \times \mathbb{R} ^{K \times r}$.

The classical remedy to remove this indeterminacy requires further (triangular-like) structure in the factors $\mat{U}$ and $\mat{V}$. For example, LU decomposition is a way forward. In contrast, we encode the invariance map (\ref{eq:symmetry_gh}) in an abstract search space by optimizing directly over a set of equivalence classes 
\begin{equation}\label{eq:equivalence-classes-balanced}
 [(\mat{U},\mat{V})] := \{ (\mat{U}\mat{M}^{-1},\mat{V}\mat{M}^{T}): \mat{M}
\in\mathrm{GL}(r) \}.
\end{equation}
The set of equivalence classes is termed as the \emph{quotient space } and is denoted by 
\begin{equation}\label{eq:quotient-balanced}
        {\mathcal{M}}_r:= \mathcal{M} /\GL{r},
\end{equation}
where the total space ${\mathcal{M}}$ is the product space $\mathbb{R}^{K
\times r} \times \mathbb{R}^{K \times r}$.

Consequently, if an element $x \in \mathcal{M}$ has the matrix characterization $(\mat{U},
\mat{V})$, then (\ref{eq:regularized_formulation}) and (\ref{eq:refining})
are of the form 
\begin{equation}
\begin{array}{lll}
\mathop {\sf minimize}\limits_{[x] \in \mathcal{M}_r} && f([x]),\\
\end{array}
\end{equation}
where $[x]=[(\mat{U},\mat{V})] $ is defined in (\ref{eq:equivalence-classes-balanced}) and $f: \mathcal{M} \rightarrow \mathbb{R}:x \mapsto f(x)$ is a \emph{smooth} function on $\mathcal{M}$, but now induced (with slight abuse of notation) on the quotient space $\mathcal{M}_r$ (\ref{eq:quotient-balanced}).

The quotient space $\mathcal{M}_r$ has the structure of a smooth \emph{Riemannian} quotient manifold of $\mathcal{M}$ by $\GL{r}$ \cite{Mishra_2014fixedrank}. The Riemannian structure conceptually transforms a rank-constrained optimization problem into an \emph{unconstrained} optimization problem over the non-linear manifold $\mathcal{M}_r$. Additionally, it allows to compute objects like gradient (of an objective function) and develop a \changeBM{Riemannian} trust-region algorithm on $\mathcal{M}_r$ that uses second-order information for faster convergence \cite{Absil_2009optimizationonManifolds}. 


\section{Optimization on Quotient Manifold}\

\begin{figure}[t]
\center
\includegraphics[scale = 0.35]{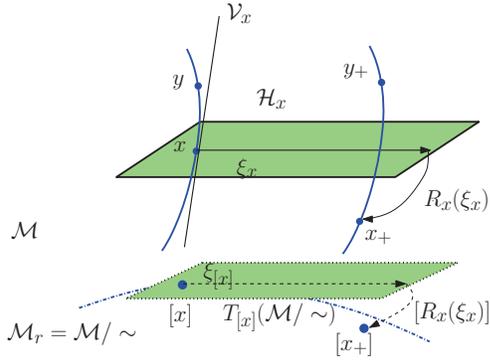}
\caption{Optimization on a quotient manifold. {The dotted lines represent abstract objects and the solid lines are their matrix representations.} The points $ {x}$ and $ y$ in the total (computational) space $ { \mathcal{M}}$ belong to the same equivalence class (shown in solid blue color) and they represent a single point $ [ {x}] :=  \{ {y} \in  {\mathcal M} :  {y} \sim  {x }\}$ in the quotient space $\mathcal{M}/\sim$. An algorithm by necessity is implemented in the computation space, but conceptually, the search is on the quotient manifold. Given a search direction $\xi_x$ at $x$, the updated point on $\mathcal{M}$ is given by the retraction mapping $R_x$.}
\label{fig:manifold_optimization}
\end{figure}

Consider an equivalence relation $\sim$ in the \emph{total} (computational)
space $ {\mathcal{M}}$. The quotient manifold ${\mathcal M}/\sim$ generated by this equivalence property consists of elements that are \emph{equivalence classes} of the form $ [ {x}] =  \{ {y} \in  {\mathcal M} :  {y} \sim  {x}\}$. Equivalently, if $[x]$ is an element in $\mathcal{M}/\sim$, then its matrix representation in $\mathcal{M}$ is $x$. Figure \ref{fig:manifold_optimization} shows a schematic viewpoint of optimization on a quotient manifold. Particularly, we need the notion of ``linearization'' of the search space, ``search'' direction and a way ``move'' on a manifold. Below we show the concrete development of these objects that allow to do develop a second-order trust-regions algorithm on manifolds.

Since the manifold $\mathcal{M}/\sim$ is an abstract space, the elements of its tangent space $T_{[x]} (\mathcal{M}/\sim)$ at $[x]$ also call for a matrix representation in the tangent space $T_x {\mathcal{M}}$ that respects the equivalence relation $\sim$. Equivalently, the matrix representation of $T_{[x]} (\mathcal{M}/\sim)$ should be restricted to the directions in the tangent space $T_{  x }  {\mathcal{M}}$ on the total space $ {\mathcal M}$ at ${  x}$ that do not induce a displacement along the equivalence class $[x]$. This is realized by decomposing $T_{  x}  {\mathcal M}$ into complementary subspaces, the \emph{vertical} and \emph{horizontal} subspaces such that $ \mathcal{V}_{  x}  \oplus \mathcal{H}_{  x} = T_{  x}  {\mathcal M}$. The vertical space $\mathcal{V}_{  x}$ is the tangent space of the equivalence class $[x]$. On the other hand, the horizontal space $\mathcal{H}_{  x}$, which is any complementary subspace to $\mathcal{V}_{  x}$ in $T_x\mathcal{M}$, provides a valid matrix representation of the abstract tangent space $T_{[x]} (\mathcal{M}/\sim)$ \cite[Section~3.5.8]{Absil_2009optimizationonManifolds}. An abstract tangent vector $\xi_{[x]} \in T_{[x]} (\mathcal{M}/\sim)$ at $ [{x}]$ has a unique element in the horizontal space $ {\xi}_{ {x}}\in\mathcal{H}_{ {x}}$ that is called its \emph{horizontal lift}. Our specific choice of the horizontal space is the subspace of $T_x \mathcal{M}$ that is the \emph{orthogonal complement} of $\mathcal{V}_{  x}$ in the sense of a Riemannian metric (an inner product).

A particular Riemannian metric on the total space ${\mathcal{M}}$ that takes into account the symmetry (\ref{eq:symmetry_gh}) imposed by the factorization model, and that is well suited to a least-squares objective \cite{mishra12a}, is
\begin{equation}\label{eq:metric_gh}
\begin{array}{lll}
{g}_{{x}}
 (    {  \xi}_{{x}} ,  {\eta}_{{x}}  )  & =  &\trace  ((\mat{V}^T\mat{V}){\xi}_{\mat
U}^T {\eta}_{\mat U})   +  \trace ( (\mat{U}^T\mat{U}) {\xi}^T_{\mat V} {\eta}_{\mat
V}  ), \\
\end{array}
\end{equation}
where ${x} = (\mat{U}, \mat{V})$ and ${\xi}_{{x}},{\eta}_{{x}} \in T_{{x}}
{\mathcal{M}}$. \changeBM{It should be noted that the tangent space $T_x\mathcal{M}$ has the matrix characterization $\mathbb{R}^{K\times r} \times \mathbb{R}^{K\times r}$. Consequently, $\eta_x$ (and similarly $\xi_x$) has the matrix representation $(\eta_{\mat U}, \eta_{\mat V}) \in \mathbb{R}^{K\times r} \times \mathbb{R}^{K\times r}$.} Motivation for the metric (\ref{eq:metric_gh}) comes from the fact that \changeBM{it is induced from a block approximation of the Hessian of a least-squares objective function}. Similar idea has also been exploited in \cite{Yuanming_2016LRMCTWC}. 

Once the metric (\ref{eq:metric_gh}) is defined on ${\mathcal M}$, \changeBM{the development of the geometric objects required for second-order optimization follow \cite{mishra12a}}. The matrix characterizations of the tangent space $T_x \mathcal{M}$, vertical space $\mathcal{V}_x$, and horizontal space $\mathcal{H}_x$ are straightforward with the expressions:
\begin{equation}\label{eq:horizontal_space_gh}
\begin{array}{lll}
T_x\mathcal{M} = \mathbb{R}^{K\times r} \times \mathbb{R}^{K\times r}\\
\mathcal{V}_{{x}} =  \{  (   -\mat{U}\mat{\Lambda} , \mat{V}\mat{\Lambda}
^T   ) : \mat{\Lambda} \in \mathbb{R}^{r\times r} \} \\
\mathcal{H}_{{x}}  = \{    ( { \zeta}_{\mat{U}} , {\zeta}_{\mat{U}}    )
:    \mat{U}^T \zeta_{\mat{U}} \mat{V}^T \mat{V}  = \mat{U}^T\mat{U} \zeta_{\mat{V}}^T
\mat{V}     , \\
\qquad \qquad \qquad \qquad  {\zeta}_\mat{U}, {\zeta}_\mat{V}  \in \mathbb{R}^{K
\times r} \}.
 \end{array}
\end{equation}

Apart from the characterization of the horizontal space, we need a linear mapping $\Pi_{{x}}: T_{{x}} {\mathcal{M}} \mapsto \mathcal{H}_{{x}}$ that projects vectors from the tangent space onto the horizontal space. Projecting an element ${\eta}_{{x}} \in T_{{x}} \mathcal{M}$ onto the horizontal space is accomplished with the operator
\begin{equation}\label{eq:projection_gh}
\Pi_{x}({\eta}_{{x}})=  
\begin{array}{ll}
( {\eta}_{\mat{U}} + \mat{U}\mat{\Lambda} , {\eta}_{\mat{V}} - \mat{V\Lambda}^T
),
\end{array}
\end{equation}
where ${\mat \Lambda} \in \mathbb{R}^{r \times r}$ is uniquely obtained by ensuring that $\Pi_{{x}}({\eta}_{{x}})$ belongs to the horizontal space characterized in (\ref{eq:horizontal_space_gh}). Finally, the expression of $\mat{\Lambda}$ is
\begin{equation*}\label{eq:Lyapunov_gh}
\begin{array}{llll}
& \mat{ U }^T ( {\eta}_{\mat{U}} + \mat{U} {\mat \Lambda}  )  \mat{V}^T \mat{V}
 = \mat{U}^T \mat{U}   (   {\eta}_{\mat{V}} - \mat{V} {\mat \Lambda} ^T
)^T \mat{V} \\


\Rightarrow & \mat{\Lambda}  =   0.5 [ \eta_{\mat{V}}^T\mat{V} (\mat{V}^
T \mat{V})^{-1}  - (\mat{U}^T\mat{U} )^{-1}\mat{U}^T\eta_{\mat{U}} ].
\end{array}
\end{equation*}

\subsection{Gradient and Hessian Computation}\label{sec:gradient_Hessian}
The choice of the metric (\ref{eq:metric_gh}) and of the horizontal space (as the orthogonal complement of $\mathcal{V}_{x}$) turns the quotient manifold $\mathcal{M}/\sim$ into a \emph{Riemannian submersion} of $({\mathcal{M}}, {g})$ \cite[Section~3.6.2]{Absil_2009optimizationonManifolds}. As shown in \cite{Absil_2009optimizationonManifolds}, this special construction allows for a convenient matrix representation of the gradient \cite[Section~3.6.2]{Absil_2009optimizationonManifolds} and the Hessian \cite[Proposition~5.3.3]{Absil_2009optimizationonManifolds} on the quotient manifold $\mathcal{M}/\sim$.

The Riemannian gradient ${\grad}_{[x]} f$ of $f$ on $\mathcal{M}/\sim$ is uniquely represented by its horizontal lift in ${\mathcal{M}}$ which has the matrix representation
\begin{equation}\label{eq:Riemannian_gradient}
\begin{array}{llll}
{\rm horizontal\ lift\ of\ }  { {\grad}_{[x]} f} \\
\qquad \quad = \grad_x  f =  (\frac{\partial f}{\partial
\mat{U}} (\mat{V}^T\mat{V})^{-1},  \frac{\partial f}{\partial
\mat{V}} (\mat{U}^T\mat{U})^{-1}),
\end{array}
\end{equation}
where $\grad_x  f$ is the gradient of $f$ in $\mathcal{M}$ and ${\partial f}/{\partial \mat{U}}$ and ${\partial f}/{\partial \mat{V}}$ are the \emph{partial derivatives} of $f$ with respect to $\mat{U}$ and $\mat{V}$, respectively.

In addition to the Riemannian gradient computation (\ref{eq:Riemannian_gradient}), we also require the directional derivative of the gradient along a search direction. This is captured by a \emph{connection} $\rc _{\xi_x} \eta_x$, which is the \emph{covariant derivative} of vector field $\eta_x$ with respect to the vector field $\xi_x$. The Riemannian connection $\rc_{\xi_{[x]}} \eta_{[x]}$ on the quotient manifold $\mathcal{M}/\sim$ is uniquely represented in terms of the Riemannian connection ${\rc}_{{\xi}_{x}} {\eta}_{x}$ in the total space ${\mathcal{M}}$ \cite[Proposition~5.3.3]{Absil_2009optimizationonManifolds} which is 
\begin{equation} \label{eq:Riemannian_connection}
{\rm horizontal\ lift\ of\ } { {\rc}_{\xi _{[x]}} {\eta _{[x]}}} = \Pi_{{x}}
({\rc}_{{\xi}_{x}} {\eta}_{x}),
\end{equation}
where $\xi_{[x]}$ and $\eta_{[x]}$ are vector fields in $\mathcal{M}/\sim$ and ${\xi}_{x}$ and ${\eta}_{x}$ are their horizontal lifts in ${\mathcal{M}}$. Here $\Pi_{x}(\cdot)$ is the projection operator defined in (\ref{eq:projection_gh}). It now remains to find out the Riemannian connection in the total space ${\mathcal{M}}$. We find the matrix expression by invoking the \emph{Koszul} formula \cite[Theorem~5.3.1]{Absil_2009optimizationonManifolds}. After a routine calculation, the final expression is \cite{mishra12a}
\begin{equation}\label{eq:connection_total_space}
\begin{array}{llll}
{\rc}_{{\xi}_x} {\eta_x}  =  \D {\eta_x}[{\xi_x}] + \left( \mat{A}_{\mat{U}},
\mat{A}_{\mat V} \right), \  {\rm where} \\

\\

\mat{A}_{\mat U}  =  {\eta}_{\mat U}\Sym( {\xi}_{\mat V} ^T \mat{V} )(\mat{V}^T\mat{V})^{-1}
    + {\xi}_{\mat U}\Sym( {\eta}_{\mat V} ^T \mat{V} )(\mat{V}^T\mat{V})^{-1}\\

\quad \qquad - \mat{U}\Sym( {\eta}_{\mat V} ^T {\xi}_{\mat V} ) (\mat{V}^T\mat{V})^{-1}
\\
\mat{A}_{\mat V}  =  {\eta}_{\mat V}\Sym( {\xi}_{\mat U} ^T \mat{U} )(\mat{U}^T\mat{U})^{-1}
    + {\xi}_{\mat V}\Sym( {\eta}_{\mat U} ^T \mat{U} )(\mat{U}^T\mat{U})^{-1}
 \\

\quad \qquad - \mat{V}\Sym( {\eta}_{\mat U} ^T \overline{\xi}_{\mat U} ) (\mat{U}^T\mat{U})^{-1}

\end{array}
\end{equation}
and $ \D {\xi}[{\eta}]$ is the Euclidean directional derivative $ \D {\xi}[{\eta}] : = \lim_{t \rightarrow 0} {({\xi}_{{x} + t {\eta}_{\bar x} } - {\xi}_{x})}/{t}$. $\Sym(\cdot)$ extracts the symmetric part of a square matrix, i.e., $\Sym(\mat{Z}) = ({\mat{Z} + \mat{Z}^T})/{2}$.

The directional derivative of the Riemannian gradient in the direction $\xi_{[x]}$ is given by the \emph{Riemannian Hessian operator} $\hess_{[x]} f [\xi _{[x]}]$ which is now directly defined in terms of the Riemannian connection $\rc$. Based on (\ref{eq:Riemannian_connection}) and (\ref{eq:connection_total_space}), the horizontal lift of the Riemannian Hessian in ${\mathcal{M}}/\sim$ has the matrix expression:
\begin{equation}\label{eq:Riemannian_hessian}
{\rm horizontal\ lift\ of\ } \hess_{[x]} f [\xi _{[x]}] = \Pi_{{x}}(  {\rc}_{{\xi}_{x}} { \grad_{ x} f}   ),
\end{equation}
where  $\xi_{[x]} \in T_{[x]} (\mathcal{M}/\sim)$ and its horizontal lift ${\xi}_{x} \in \mathcal{H}_{{x}} $. $\Pi_x(\cdot)$ is the projection operator defined in (\ref{eq:projection_gh}).



\subsection{Retraction}
An iterative optimization algorithm involves computing a  search direction (e.g., negative gradient) and then ``moving in that direction''. The default option on a Riemannian manifold is to move along geodesics, leading to the definition of the \emph{exponential map}. Because the calculation of the exponential map can be computationally demanding, it is customary in the context of manifold optimization to relax the constraint of moving along geodesics. \changeBM{To this end, we define} {\it retraction} $R_{ x}: \mathcal{H}_{ x}  \rightarrow \mathcal{M}: \xi_x \mapsto R_x(\xi_x)$ \cite[Definition~4.1.1]{Absil_2009optimizationonManifolds}. A natural update on the manifold $\mathcal{M}$ is, therefore, based on the update formula $x_+ = R_x(\xi_x)$, i.e., defined as
\begin{equation}\label{eq:retraction_gh}
\begin{array}{lll}
R_{\mat{U}} ({\xi}_{\mat U}) = \mat{U} + {\xi}_{\mat U} \\
R_{\mat{V}} ({\xi}_{\mat V}) = \mat{V} + {\xi}_{\mat V}, \\
\end{array}
\end{equation}
where ${\xi}_{ x} =(\xi_{\mat U}, \xi_{\mat V}) \in \mathcal{H}_{ x}$ is a search direction and ${x}_+ \in {\mathcal M}$. \changeBM{It translates into the update  $[x_+] = [R_x(\xi_x)]$ on $\mathcal{M}/\sim$.}

\subsection{Riemannian Trust-Region Algorithm}\label{sec:trust_regions}
Analogous to trust-region algorithms in the Euclidean space \cite[Chapter~4]{Wright_2006numericalopt}, trust-region algorithms on a Riemannian quotient manifold with guaranteed superlinear rate convergence and global convergence have been proposed in \cite[Chapter~7]{Absil_2009optimizationonManifolds}. At each iteration we solve the \emph{trust-region sub-problem} on the quotient manifold $\mathcal{M}/\sim$. The trust-region sub-problem is formulated as the minimization of the \emph{locally-quadratic} model of the objective function. \changeBM{The concrete matrix characterizations} of Riemannian gradient (\ref{eq:Riemannian_gradient}), Riemannian Hessian (\ref{eq:Riemannian_hessian}), projection operator (\ref{eq:projection_gh}), and retraction (\ref{eq:retraction_gh}) allow to use an \emph{off-the-shelf} trust-region implementation on manifolds, e.g., in Manopt \cite{manopt}.

\section{Simulation Results}
In this section, we compare the proposed Riemannian optimization algorithm in Table \ref{tab:overall_algorithm} with with the alternating minimization algorithm based on (\ref{l1norm_factorized}) for the sparse and low-rank optimization problem $\mathscr{P}$. For the alternating minimization algorithm, we need to solve a sequence of subproblems with non-smooth $\ell_1$-norm objective and an affine constraint (i.e., linear programming problem) \changeBM{for which we use} CVX \cite{cvx}. \changeBM{The maximum number of iterations of the proposed alternating minimization algorithm is set to be $50$.} \changeBM{For the proposed Riemannian algorithm, we set $\epsilon$ to a high value of $0.01$. A good choice of $\rho$ is $0.001$ and is obtained by cross-validation. The Riemannian algorithm in Table \ref{tab:overall_algorithm} is implemented in Manopt \cite{manopt}. The maximum number of trust-region iterations is set to $100$. \changeBM{The Matlab codes are available at https://bamdevmishra.com/codes/indexcoding.}}

\begin{figure}[t]
  \centering
  \includegraphics[scale=0.8]{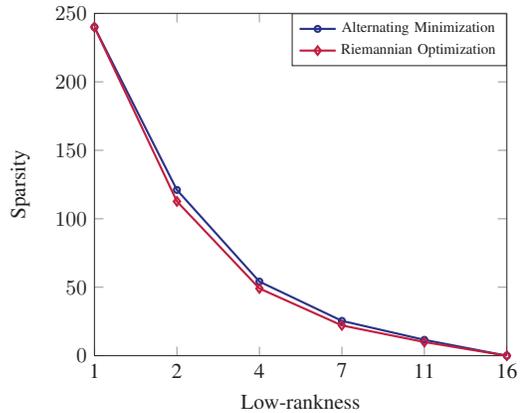}
 \caption{\changeYS{Sparsity  and  low-rankness tradeoff in matrix $\bm{X}$, where sparsity is given by $(\|\bm{X}\|_0-K)$ and the low-rankness is given by ${\rm{rank}}(\bm{X})$.}}
 \label{slopt}
\end{figure}

Consider a sparse and low-rank optimization problem $\mathscr{P}$ with $K=16$. \changeBM{(We consider a smaller size instance as CVX is too computationally expensive to run larger ones.)} The achievable normalized data rate equals $1/{\rm{rank}}(\bm{X})$, and the amount of normalized side information equals $(\|\bm{X}\|_0-K)$, which measures the cache size. Therefore, the sparsity and \changeYS{low-rankness} tradeoff in Figure {\ref{slopt}} reveals the tradeoff between the amount of side information and the achievable data rate in the corresponding index coding problem. Furthermore, Figure {\ref{slopt}} demonstrates that, by encoding the second-order information in the algorithm design, the trust-region Riemannian algorithm can achieve sparser solutions than the alternating minimization algorithm.

\section{Conclusion}
In this paper, we proposed a new sparse and low-rank optimization modeling framework to characterize the tradeoff between the amount of the side information and the achievable data rate by revealing the sparsity and low-rankness tradeoff in  the \changeYS{modeling matrix}.  A trust-region Riemannian optimization algorithm was proposed to improve the performance by encoding the second-order information, as well as the quotient manifold geometry of the fixed-rank matrices in the search space. This is achieved by relaxing the $\ell_0$-norm as a smooth $\ell_1$-norm surrogate and regularizing the affine constraint with least-squares objective. Simulation results revealed the fundamental tradeoff between the amount of side information and the achievable data rate in index coding problem. Our framework is useful for important system design problems, e.g., cache size allocation. A promising and interesting future research direction is theoretically characterizing the fundamental tradeoffs between storage size and the achievable data rate, i.e., the sparsity and low-rankness tradeoff in the proposed \changeYS{modeling matrix}.          

\bibliographystyle{ieeetr}

\end{document}